\documentclass[aps,prl,twocolumn,showpacs]{revtex4-1}

\usepackage{graphicx}

\usepackage{amssymb}

\begin{document}

\title{ Threefold Reduction in Thermal Conductivity of Vycor Glass Due to Adsorption of Liquid $^4$He }

\author{Z. G. Cheng}
\author{M. H. W. Chan} \email{chan@phys.psu.edu}

\affiliation{Department of Physics, The Pennsylvania State University, University Park, PA 16802, USA}

\date{\today}

\begin{abstract}
We report thermal conductivity measurements of porous Vycor glass when it is empty and when the pores are filled with helium between 0.06 and 0.5 K. The filling of liquid $^3$He and liquid $^4$He inside the Vycor pores brings about respectively two and three fold \emph{reduction} of the thermal conductivity as compared with empty Vycor. This dramatic reduction of thermal conductivity, not seen with solid $^3$He and $^4$He in the pores, is the consequence of hydrodynamic sound modes in liquid helium that greatly facilitate the quantum tunneling of the two-level systems (TLS) in Vycor and enhance the scattering of the thermal phonons in the silica network.
\end{abstract}


\maketitle
The atoms or molecules in a disordered glassy material have multiple potential energy minima in their spatial configurations. At sufficiently low temperature, typically below 1 K, the thermodynamic properties of the material are dominated by quantum tunneling between two adjacent accessible energy levels. The predictions of this two-level system (TLS) model \cite{AndersonPW, Phillips1, Jackle}, including a specific heat $C$ scaling with temperature $T$ and a thermal conductivity $\kappa$ scaling with $T^2$, have been confirmed by experiments \cite{Zeller, Stephens1,  Lasjaunias}.

Zeller \emph{et al.} found that the thermal conductivity of vitreous silica below 1 K is much lower than its crystalline counterpart, quartz, and scales with $T^2$ instead of $T^3$ \cite{Zeller}. Heat is conducted in solid by thermally activated phonons and the thermal conductivity is given by ${\kappa}=(1/3)c_{phonon}\bar{u}l$, where $c_{phonon}$ is the contribution to the specific heat per volume by the phonons, $\bar{u}$ the average sound velocity and $l$ the phonon mean free path. Normally $c_{phonon}{\sim}T^3$ and $\bar{u}$ is weakly dependent on $T$. At low temperatures, phonons propagating in crystalline solid like quartz are primarily scattered at the defects and boundaries of the sample. Therefore $l$ is temperature independent and ${\kappa}{\sim}T^3$. For amorphous solid like vitreous silica, phonons are dominantly scattered by TLS that involve the coordinated motion and rotation of SiO$_4$ tetrahedra groups. The physical dimension of each group, based on simulation and neutron scattering studies, is more than 1 nm \cite{Guttman, Buchenau1, Buchenau2, Phillips2}. TLS scatter phonons by absorbing resonant phonons to tunnel to the excited state and re-emitting phonons to tunnel back to the ground state with a time constant $\tau$. The re-emitted or scattered phonons are generally propagating in a different direction from the incident phonons. TLS-phonon scattering results in an $l$ that scales with $T^{-1}$ \cite{Jackle}. As a result ${\kappa}{\sim}T^2$.  If there are mechanisms to enhance the TLS tunneling rate ${\tau}^{-1}$, more phonons will be scattered per unit time and $\kappa$ will be reduced further. We show in this paper that the infusion of liquid helium into Vycor pores dramatically reduces $\kappa$.

In addition to the TLS-phonon scattering, phonons are also scattered by the porous structure when propagating in a porous glass like Vycor. Nitrogen adsorption isotherm measurements at $T=77~\mathrm{K}$ show that Vycor has a porosity of 28\% and a specific surface area of about $100~\mathrm{m^2/g}$ \cite{Wallacher, Levitz}. Transmission electron microscopy (TEM) revealed the pore space to be a network of multiply connected cylindrical channels \cite{Levitz}. The diameter of the channel and the thickness of the silica ``walls'', based on TEM and adsorption studies, are both found to be $\sim$7 nm \cite{Wallacher, Levitz}. The spatial correlation length of the silica structure (also the pore structure) in Vycor, $\xi$, is found to be distributed between 5 to 60 nm by small angle X-ray and neutron scattering and TEM studies \cite{Levitz, Schaefer, Wiltzius}. Therefore, phonons with wavelengths near or shorter than 60 nm are expected to be strongly scattered by the Vycor porous structure because of the acoustic mismatch at the silica-pore interface. Phonons with wavelength much longer than 60 nm, on the other hand, will be insensitive to the porous structure and  scattered solely by TLS. The spectrum of thermal phonons obeys the Planck$^{\prime}$s distribution which shows a peak at ${\lambda}_{dom}=hu/4.2k_{B}T$, the characteristic or the dominant wavelength \cite{Zeller} and the half maxima of the distribution at $0.5{\lambda}_{dom}$ and $3{\lambda}_{dom}$. Here $h$ is the Planck$^{\prime}$s constant, $u$ the sound velocity and $k_{B}$ the Boltzmann constant. Since the temperature range of this experiment is between 0.06 and 0.5 K, the characteristic frequency of thermal phonons, $f_{dom}=4.2k_{B}T/h$ is between 5 and 40 GHz. Taking the transverse sound velocity in Vycor, $u_{Vycor}$ to be 2200 m/s \cite{Mulders}, ${\lambda}_{dom}$ is found to be 420, 170, 50 nm at $T=0.06,0.15,0.5 ~\mathrm{K}$, respectively. At low temperature when ${\lambda}_{dom}$ is much longer than $\xi$, phonons are mainly scattered by TLS. As a consequence, $\kappa$ should scale with $T^2$. With increasing temperature, ${\lambda}_{dom}$ decreases towards $\xi$ and the scattering effect by the porous structure becomes prominent. This will reduce $\kappa$ and result in sub-quadratic temperature dependence. This behavior has been experimentally confirmed \cite{Stephens2, Tait, Hsieh}.

In this Letter, we present the thermal conductivity of Vycor infused by solid helium, liquid helium and helium films. Remarkably and counter-intuitively, a significant \emph{reduction} in $\kappa$, up to a factor of three, is observed when Vycor is infused with superfluid $^4$He films, liquid $^4$He and liquid $^3$He. 

\begin{figure}[t]
    \centerline{\includegraphics[width=1\columnwidth]{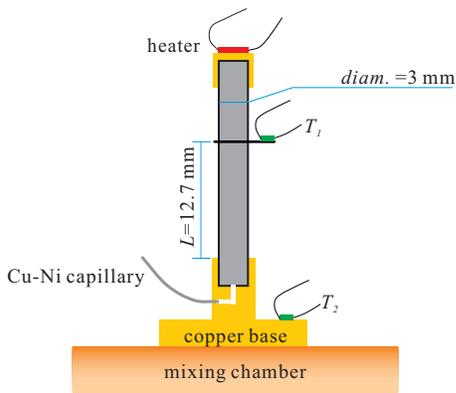}}
    \caption{\label{Fig1}Experimental setup of thermal conductivity measurements.}
\end{figure}

\begin{figure}[b]
    \centerline{\includegraphics[width=1\columnwidth]{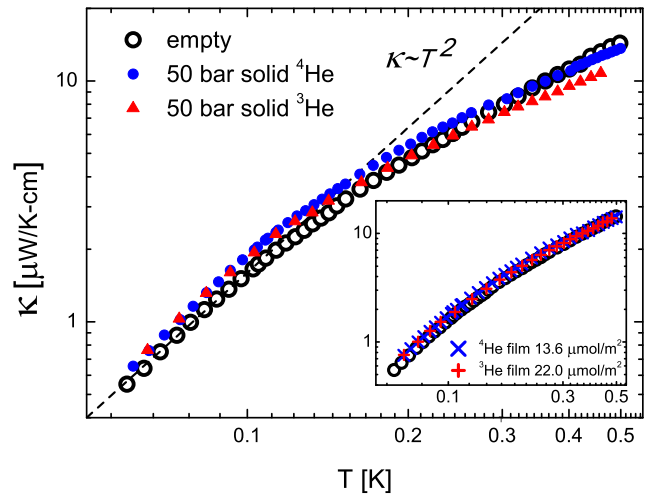}}
    \caption{\label{Fig2}Thermal conductivity of empty Vycor and Vycor with solid $^3$He and $^4$He at 50 bar. $\kappa$ of solid $^4$He sample at 47 bar, not shown in the figure, was found to lie between the two data sets shown in the figure over the entire temperature range. Inset shows thermal conductivity of Vycor with inert helium films.}
\end{figure}

\begin{figure}[t]
    \centerline{\includegraphics[width=1\columnwidth]{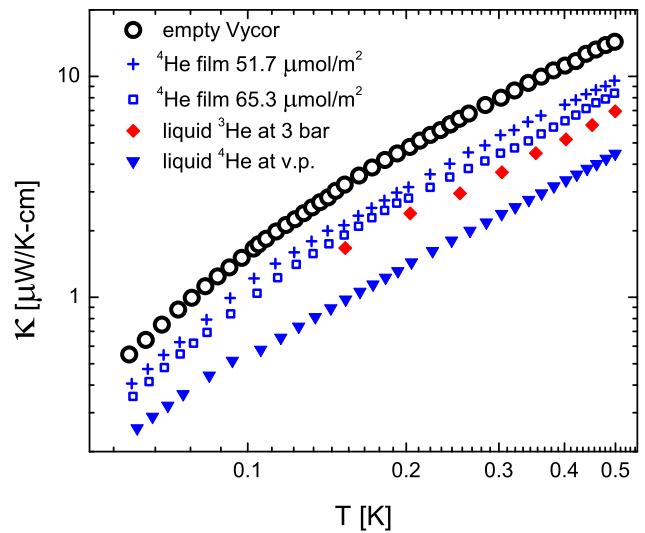}}
    \caption{\label{Fig3}Thermal conductivity of empty Vycor and Vycor with liquid $^4$He, liquid $^3$He and superfluid $^4$He films with superfluid transition temperatures well above 1 K.}
\end{figure}

Fig. \ref{Fig1} shows our experimental setup. The Vycor rod is 3 mm in diameter and 22.8 mm long. It is secured mechanically and thermally into a copper base with epoxy resin (Stycast 2850FT) and attached to the mixing chamber of a dilution refrigerator. Outside the copper base, the Vycor rod is sealed only by a thin layer of epoxy resin less than 0.07 mm thick painted on its outer surface. The thin epoxy layer, impregnating the pores on the surface, is able to hold pressure up to at least 85 bar below 4 K but does not measurably contribute to the thermal conductance of the sample. Such an experimental configuration was recently used in studies of supersolidity in solid helium \cite{Ray, Kim}. Helium is introduced into Vycor through the copper base with a thin Cu-Ni capillary. A heater is attached at the top of the Vycor rod. A germanium thermometer is secured 12.7 mm away from the copper base reading $T_1$ and another thermometer is attached directly onto the copper base reading $T_2$. The copper base is at a uniform temperature because the thermal conductivity of copper is $10^5$ times higher than Vycor \cite{Lounasmaa}. Thermal conductivity of the Vycor rod is given by ${\kappa}=L{\dot{Q}}/A{\Delta}T$, where $L$ is the distance between the germanium thermometer and the copper base, $A$ is cross section area of Vycor, $\dot{Q}$ is the steady state heat power and ${\Delta}T=T_{1}-T_{2}$. Measurements are made by imposing a dc power between 0.1 and 1.6 nW to maintain ${\Delta}T$ small but experimentally significant (typically 2 mK). The measured $\kappa$ of empty Vycor from 0.06 to 0.5 K is shown in both Fig. \ref{Fig2} and Fig. \ref{Fig3}. As noted above, the deviation below the $T^2$ dependence for $T>0.15 ~\mathrm{K}$ is clearly evident. Our results are consistent with previous experiments within a few percent \cite{Stephens2, Tait, Hsieh}. 

After the measurements of empty Vycor, solid, liquid and adsorbed films of  both helium isotopes were introduced into the Vycor pores. For ease of description, these samples will be identified as solid, liquid and film samples. For clarity, we plot $\kappa$ of the solid and atomically thin film samples in Fig. \ref{Fig2} and $\kappa$ of liquid and superfluid film samples in Fig. \ref{Fig3}. When Vycor is infused with liquid or solid helium, the samples can be considered as composites consisting of the two intertwining (silica and helium) networks. However, the heat is still conducted primarily by the silica network. This is the case because the sound velocities (and hence phonon wavelengths) in liquid and solid helium are $\sim$10 times smaller than that in silica. Therefore the phonon wavelengths in helium are always shorter than 60 nm for $T>0.06 ~\mathrm{K}$. As a result, phonon propagation along the helium network, if exists, is scattered by the porous structure much more strongly than that along the silica network.

As shown in the inset of Fig. \ref{Fig2}, $\kappa$ of the atomically thin $^4$He and $^3$He film samples are slightly ($\sim$3\%) higher than that of the empty Vycor over the entire temperature range. The helium atoms in these thin films are immobile and tightly bound to the random silica surface with a typical adsorption potential higher than 25 K for the $^4$He film (surface coverage $n_{4}=13.6 ~\mathrm{{\mu}mol/m^2}$) and 3 K for the $^3$He film ($n_{3}=22 ~\mathrm{{\mu}mol/m^2}$) \cite{Sabisky}. The $^4$He film does not show superfluidity even at $T=0 ~\mathrm{K}$. In our temperature range of interest, the adsorbed atoms are therefore “inert” and merely serve to thicken the silica network. These tightly bound atoms soften the acoustic mismatch at the silica-pore interface and lead to the slight enhancement of $\kappa$ as compared with the empty Vycor. A more noticeable enhancement of $\kappa$ is observed for the solid helium samples in the low temperature limit. However, as $T$ is increased above 0.2 and 0.36 K,  $\kappa$ of the solid $^3$He and $^4$He samples respectively crossover from being higher to being lower than the empty Vycor value. We attribute the reduction of $\kappa$ to the coupling between the phonons in solid helium and the TLS tunneling in silica. For empty Vycor, phonons absorbed by TLS in silica can only be re-emitted to the silica structure. But for Vycor filled with solid helium, TLS have a second channel to release energy by exciting phonons in solid helium via modulating the van der Waals interaction at the silica-helium interface \cite{Kinder, Beamish1}. The TLS-helium coupling facilitates the TLS tunneling and enhances the tunneling rate ${\tau}^{-1}$. As a result TLS-phonon scattering in silica becomes stronger and $\kappa$ is reduced. However phonon excitation in solid helium is strongly suppressed in the low temperature limit when the phonon wavelength ${\lambda}_{He}$ exceeds the pore dimension. This is why an enhancement of $\kappa$ is seen in the solid samples for $T<0.2 ~\mathrm{K}$ due to the softening of acoustic mismatch across the silica-solid helium interface. With increasing temperature, ${\lambda}_{He}$ decreases to become comparable and then shorter than the pore dimension thus opening the TLS-helium coupling channel and resulting in a reduction of $\kappa$. The difference in the “crossover” temperatures of the solid $^3$He (0.2 K) and solid $^4$He (0.36 K) samples reflects the different transverse sound velocities $u_t$ and hence ${\lambda}_{He}$ of the two solid helium samples. $u_t$ of 50 bar solid $^3$He is 180 m/s. This translates to a ${\lambda}_{He}$ of $\sim$10 nm to match the pore dimension of Vycor near 0.2 K. In comparison, $u_t$ of 50 bar solid $^4$He is 270 m/s which translates to a ${\lambda}_{He}$ of  $\sim$10 nm near 0.36 K. The modest difference in $\kappa$ ($<20\%$) between the solid samples and the empty Vycor over the entire temperature range of the experiment confirms the reasoning stated above that heat conduction is primarily along the silica network and perturbed by the presence of helium.

Although the density and (first) sound velocity of liquid helium are similar as those of solid helium, the measured $\kappa$ of the liquid samples are dramatically different. Instead of a modest change, $\kappa$ of full pore liquid $^4$He sample is reduced threefold as compared with empty Vycor over the entire temperature range as shown in Fig. \ref{Fig3}. A smaller but still sizable (a factor of 1.5) reduction of $\kappa$ is also seen in unsaturated superfluid $^4$He film samples ($n_{4}=51.7, ~65.3 ~\mathrm{{\mu}mol/m^{2}}$). Interestingly, a two-fold reduction of $\kappa$ is found when the pores are filled with non-superfluid liquid $^3$He. This suggests the dramatic reduction of $\kappa$ is a consequence of fluidity and not superfluidity of liquid helium.

The fluidity of the liquid helium enables hydrodynamic sound modes to be excited in the porous structure. Biot showed that such excitation requires the viscous penetration depth of the liquid ${\delta}=\sqrt{{\eta}/{\pi}{\rho}f}$ to be smaller than the pore radius $D/2$, so that a fraction of liquid can be decoupled and in relative motion with respect to the solid (in our case the silica) matrix \cite{Biot1, Biot2}. Here $\eta$ and $\rho$ are viscosity and density of the liquid, $f$ is the sound frequency. Similar to phonons in solid helium, the hydrodynamic sounds are also excited by TLS via the modulation of van der Waals interaction at the silica-helium interface. But in contrast to phonons in solid helium in the pores which must satisfy the boundary condition that their wavelengths should be comparable or shorter than the pore dimensions, there is no such wavelength (and hence frequency) restriction for the hydrodynamic sound modes because helium atoms are mobile. The sound modes conform to and also extend to the entire multiply connected pore space. The frequency spectrum of this hydrodynamic sound is continuous and is given by the Planck$^{\prime}$s distribution of thermal phonons in silica with the additional Biot condition: $f>f_{c}=4{\eta}/{\pi}{\rho}D^2$. Therefore the hydrodynamic sound is much more efficient in enhancing ${\tau}^{-1}$ of TLS in the silica over the entire temperature range as compared with phonons in solid helium. This is the key reason for the dramatic reduction of $\kappa$ in the liquid samples. The hydrodynamic sound is often named as “slow wave” or “slow sound” \cite{Plona, Johnson1, Johnson2, Beamish2, Kwon} because its velocity is slower than the sound velocity in the solid matrix. Johnson pointed out that the slow sound in superfluid $^4$He confined in porous media is the well-known fourth sound \cite{Johnson1}. Fourth sound can be excited at any finite frequency within the Planck$^{\prime}$s distribution because the zero viscosity of superfluid gives $f_{c}=0$. In the case of unsaturated superfluid films where viscosity is also zero, third sound plays the role of slow sound in enhancing ${\tau}^{-1}$. The fourth and third sounds themselves do not contribute to thermal conductivity because superfluid carries no entropy. 

Slow sound can also be excited in liquid $^3$He with a finite viscosity as long as $f>f_c$. The density of liquid $^3$He at 3 bar (pressure of our sample) is $0.089 ~\mathrm{g/cm^3}$. The viscosity ranges from 150 to 50 $\mu$Poise between 0.15 K and 0.5 K \cite{Black}. As a result $f_c$ ranges from 4.4 to 1.5 GHz, substantially lower than the characteristic frequency of the slow sound $f$, equivalent to $f_{dom}$, ranging from 12 to 40 GHz in the same temperature range. Therefore, there is a significant fraction of liquid $^3$He in the pores of Vycor that supports the slow sound to enhance ${\tau}^{-1}$. The slow sound in liquid $^3$He is expected to vanish below 0.07 K when the viscosity of liquid $^3$He is large enough to completely lock the liquid to the silica matrix. Unfortunately, the large heat capacity of $^3$He and the low thermal conductivity of the Vycor prevent us from cooling the composite below 0.15 K.

The TLS tunneling rate in an amorphous system has been shown to be \cite{Jackle, Kinder} 
\begin{equation}
\tau^{-1}=\frac{M^{2}E^{3}}{2{\pi}{\hbar}^{4}{\rho}u^{5}}\coth{\frac{E}{2k_{B}T}} \label{eq1}
\end{equation}
Here $E$ is TLS energy separation, $\rho$ and $u$ are the density and sound velocity of the medium where TLS release energy, and $M$ is the deformation potential that characterizes the coupling strength between TLS tunneling and the resultant strain in the medium. For empty Vycor, the deformation occurs at Si-O bonds and $M_{Si-O}$ is on the order of  1 eV \cite{Jackle}. In this case the TLS tunneling rate is labeled as ${\tau}_{Si-O}^{-1}$. With liquid helium in the pores, the TLS on the pore surface also couple with helium via modulating the silica-helium van der Waals interaction and $M_{SiO_{2}-He}$ is $\sim$2 meV \cite{Kinder}. In this case the total TLS tunneling rate is given by ${\tau}^{-1}={\tau}_{Si-O}^{-1}+{\tau}_{SiO_{2}-He}^{-1}$. Since the density and sound velocity of liquid helium are $\sim$10 times smaller than those of silica, Eq. (\ref{eq1}) shows that ${\tau}_{SiO_{2}-He}^{-1}$ is larger than ${\tau}_{Si-O}^{-1}$ by a factor of 4. As a result, ${\tau}^{-1}$ is enhanced by a factor of 5. Since the physical dimension of the SiO$_4$ tetrahedra groups that make up the TLS is more than 1 nm and the thickness of silica ``walls'' is $\sim$7 nm, at least 30\% of the TLS reside on the pore surface. The average ${\tau}^{-1}$ of the entire sample is therefore enhanced at least by 2.2 times, close to the three-fold reduction in $\kappa$ found in full pore liquid $^4$He sample. This agreement supports the model we are proposing for the observed reduction of $\kappa$. Superfluid films cause less reduction of $\kappa$ than full pore liquid $^4$He because less helium are coupled to TLS. In the liquid $^3$He sample, the slow sound propagates with finite damping because of the finite viscosity. The damping may dissipate energy to the silica network in the form of phonons. This may explain that liquid $^3$He causes less reduction in $\kappa$ than liquid $^4$He. 

Beamish \emph{et al.} \cite{Beamish1} and Mulders \emph{et al.} \cite{Mulders} also reported an enhancement of TLS tunneling rate in Vycor due to the adsorption of liquid and solid helium in ultrasound experiments between 0.08 and 5 K. However, these experiments were not probing thermal phonons since the ultrasound frequencies correspond to characteristic temperature  between 0.3 and 10 mK. Schubert \emph{et al.} observed that the transmission of 25-GHz phonons from quartz to amorphous paraffin film is enhanced by the adsorption of liquid helium film on the paraffin at 1.8 K \cite{Schubert}. They attributed this to the enhancement of TLS tunneling in the amorphous paraffin film.

In summary, we observe a dramatic reduction of thermal conductivity of Vycor when the pores are filled with liquid helium and superfluid films. The reduction is a consequence of hydrodynamic slow sounds in liquid $^4$He, superfluid $^4$He films and liquid $^3$He that facilitate the TLS tunneling in silica and  enhance the TLS-phonon scattering. 

\begin{acknowledgments}
We thank John Beamish, Norbert Mulders and Robert Hallock for useful discussions. We also thank Stefan Omelchenko and Samhita Banavar for assistance in the early stages of the experiment. This research was supported by NSF Grant No. DMR1103159.
\end{acknowledgments}

\bibliography{TC1}

\end{document}